\documentclass[%
 twocolumn,
 amsmath,amssymb,
 aps,
 bibliography, 
]{revtex4-1}

\usepackage{graphicx}
\usepackage{dcolumn}
\usepackage{bm}
\usepackage{url}
\usepackage[dvipdfmx]{hyperref}%
\usepackage{enumerate}

\begin{document}

\title{Gauge coupling unification with extra Higgs doublets}%

\author{Junpei Harada}
 \email{jharada@hoku-iryo-u.ac.jp}
\affiliation{%
 Research Center for Higher Education, Health Sciences University of Hokkaido, Japan
}%

\date{2 May, 2016}

\begin{abstract} 
Gauge coupling unification is studied within the framework where there are extra Higgs doublets and $E_6$ exotic fields. Supersymmetric models and nonsupersymmetric models are investigated, and a catalog of models with gauge coupling unification is presented.
 \end{abstract}
 
\maketitle

\section{Introduction}
The Standard Model of particle physics contains a single Higgs doublet field, and the recent discovery of a scalar boson at the CERN Large Hadron Collider (LHC)~\cite{Aad:2012tfa, Chatrchyan:2012xdj} indicates that the Higgs boson really exists. However, with the current experimental data, it is impossible to determine how many Higgs doublet(s) should exist, and extending the Higgs sector of the Standard Model is indeed one of the attractive candidates for physics beyond the Standard Model. 

A two-Higgs-doublet model (2HDM) (for a review, see ref.~\cite{Branco:2011iw}) has been most extensively studied in the past, and a multi-Higgs-doublet model (MHDM)~\cite{Grossman:1994jb} in which there are more than two Higgs doublets has also been investigated. Particularly, the special case that there are three-Higgs generations~\cite{Kilian:2006hh, Hartmann:2014ppa} is interesting, it is motivated by a matter-Higgs unification in $E_6$, where matter superfields and the doublet Higgs superfields are unified in a single gauge multiplet of $E_6$. 

Extra Higgs doublets contribute to the renormalization running of gauge coupling constants of the Standard Model, and therefore gauge coupling unification of the minimal supersymmetric Standard Model (MSSM) is modified. With three Higgs generations, gauge coupling unification has been investigated within the framework of the exceptional supersymmetric Standard Model ($E_6$SSM)~\cite{King:2007uj}, and of Pati-Salam unification models~\cite{Hartmann:2014fya}. Very recently, gauge coupling unification has been studied in supersymmetric $E_6$ models~\cite{Cho:2016afr}. 

In this paper we study gauge coupling unification within the framework where there are extra Higgs doublets and/or $E_6$ exotic fields. We investigate both supersymmetric and nonsupersymmetric models, and give a catalog of models in which gauge coupling constants of the Standard Model are unified.

Particularly, we show that in supersymmetric three-Higgs-generation model gauge coupling constants are unified, if two generations of $E_6$ exotic fields are light and one generation is decoupled. This special case is interesting, because the lifetime of proton is estimated as the experimentally observable level in the near future. The value of gauge coupling constant at the grand unification scale is $\alpha^{-1} \simeq 14$, and it enhances the decay rate of nucleus. For a typical value of mass of the grand unified gauge boson, the lifetime is estimated as $\tau/B(p \rightarrow e^+\pi^0) \simeq 4 \times 10^{34}$ years that is just above the current Super-Kamiokande limit $\tau/B(p\rightarrow e^+ \pi^0) > 1.29 \times 10^{34}$ years~\cite{Nishino:2012bnw}. 

We also demonstrate that even in nonsupersymmetric theories gauge coupling constants of the Standard Model are unified if there are eight Higgs doublets rather than a single Higgs doublet. In this model the grand unification scale is approximately given by $5 \times 10^{13}$ GeV. 

\section{Matter-parity in $E_6$}
Grand unified theories with $E_6$ gauge group have many theoretical advantages---anomaly free, the largest exceptional Lie group that has chiral representation, a possible unification of quarks, charged leptons, neutrinos and Higgs doublets---and therefore much attention has been paid in the literature~\cite{Gursey:1975ki, Achiman:1978vg, Ruegg:1979fr, Barbieri:1981yy, Gursey:1981kf, Robinett:1982gw, Sen:1985af, Robinett:1985dz, Bando:1999km, Bando:2000gs, Harada:2003sb,Maekawa:2003ka, Stech:2003sb, Frank:2004vg, Stech:2008wd, Kawase:2010na, Stech:2010tv, Chen:2010tg, DiLuzio:2011my, Kawamura:2013rj, Huang:2014zba, and:2015uya, Cho:2016afr}.

We begin with the mention of matter-parity $P_{\rm M}\equiv (-1)^{3(B-L)}$ in $E_{6}$, where $B$ is baryon number and $L$ is lepton number. All the matter fields---quarks, charged leptons, neutrinos, and their supersymmetric particles---have matter-parity of $-1$, and the Higgs fields and their supersymmetric particles have matter-parity of $+1$. In contrast to $SU(5)$ or $SO(10)$, matter-parity {\it is not} orthogonal to $E_6$ as shown below, and it leads a matter-Higgs unification. In $SU(5)$ or $SO(10)$ models, matter-parity {\it is} orthogonal to $SU(5)$ or $SO(10)$, and therefore it is impossible to unify matter and the Higgs superfields in a single gauge multiplet. 

The fundamental representation $\bm{27}$ of $E_6$ is decomposed as follows, 
\begin{align}
\begin{array}{ccccccll}
	\bm{27} &= & \bm{16}_1 & \oplus & \bm{10}_{-2} & \oplus & \bm{1}_{4}, \\
			&= & \overbrace{\bm{10}_{-1} \oplus \bm{5}_{3}^* \oplus \bm{1}_{-5}} & \oplus & \overbrace{\bm{5}_{2} \oplus \bm{5}_{-2}^*} & \oplus & \bm{1}_0, & \label{eq:27E6}
\end{array}
\end{align}
where the decomposition is shown for $E_6 \supset SO(10) \otimes U(1)_{V^\prime}$ in the first line and subscript is $U(1)_{V^\prime}$ charge, and for $SO(10) \supset SU(5) \otimes U(1)_V$ in the second line and subscript is $U(1)_V$ charge. An another notation has also been used in the literature---$U(1)_\psi$ for $U(1)_{V^\prime}$, $U(1)_\chi$ for $U(1)_V$---this is just a notation issue, and in this paper we use the notation of eq.~\eqref{eq:27E6}. Furthermore, $\bm{10}$, $\bm{5}^*$, $\bm{1}$ representations of $SU(5)$ are decomposed as follows, 
\begin{align}
\begin{array}{ccccccc}
	\bm{10} &=&	(\bm{3}, \bm{2})_{\frac{1}{6}} &\oplus& (\bm{3}^*, \bm{1})_{-\frac{2}{3}}& \oplus & (\bm{1}, \bm{1})_1, \\
	\bm{5}^* &=& (\bm{3}^*, \bm{1})_{\frac{1}{3}}& \oplus &(\bm{1}, \bm{2})_{-\frac{1}{2}}, \\
	\bm{1} &=& (\bm{1}, \bm{1})_0,
\end{array}
\end{align}
where decomposition is shown for $SU(5) \supset SU(3)_C \otimes SU(2)_L \otimes U(1)_Z$, the subscript is $U(1)_Z$ charge, and two numbers in parentheses indicate $SU(3)_C$ and $SU(2)_L$ representations, respectively. 

If $E_6$ contains $U(1)_{B-L}$ as a subgroup, it means that 
\begin{align}
	U(1)_Z \otimes U(1)_V \otimes U(1)_{V^\prime} \supset U(1)_{B-L}. 
\end{align}
It is known that there are three assignments that are consistent with the Standard Model~\cite{Harada:2003sb},
\begin{align}
	1)  B-L &=-\frac{1}{5}(V - 4Z), \ \qquad \qquad SO(10) \supset U(1)_{B-L} \label{eq:B-L1}\\
	2) B-L &= +\frac{1}{20}(16 Z + V + 5V^\prime),  \ SO(10) \not\supset U(1)_{B-L} \label{eq:B-L2}\\
	3)  B-L &= -\frac{1}{20}(8Z + 3V - 5V^\prime), \ SO(10) \not\supset U(1)_{B-L} \label{eq:B-L3}
\end{align}
where $Z$, $V$ and $V^\prime$ are the charges of $U(1)_Z$, $U(1)_V$ and $U(1)_{V^\prime}$, respectively. These three assignments are related by $SU(2)$~\cite{Harada:2003sb}, which is a subgroup of $SU(3)_R$ where $E_6 \supset SU(3)_C \otimes SU(3)_L \otimes SU(3)_R$. Although matter-parity $P_{\rm M}$ depends on the choice of these three assignments, in any cases matter-parity is not orthogonal to $E_6$ as follows.

The first assignment given by eq.~\eqref{eq:B-L1} is standard in the meaning that it has been most widely investigated in the literature. In this case, matter-parity can be written as
\begin{align}
		\bm{27}  =  \bm{16}_{-}  \oplus  \bm{10}_{+}  \oplus  \bm{1}_{+} \qquad E_6 \supset SO(10), \label{eq:matter-parity_1st}
\end{align}
where subscript indicates matter-parity. Thus, a spinor representation $\bm{16}$ of $SO(10)$ has matter-parity of $-1$, $\bm{10}$ and $\bm{1}$ have matter-parity of $+1$. Therefore, matter superfields with matter-parity of $-1$, and the Higgs superfields with matter-parity of $+1$, can be unified in a single gauge multiplet of $E_6$.

Similar relations are realized for the second and the third assignments. In the case of the second assignment defined by eq.~\eqref{eq:B-L2}, however, matter-parity is not orthogonal to $SO(10)$. In this case, matter-parity is orthogonal to $SU(5)$, and then matter-parity can be written as follows, 
\begin{align}
	\begin{array}{rccccccl}
	\bm{27} &=& \bm{16} &\oplus& \bm{10} &\oplus& \bm{1}  \\
	&=& \overbrace{\bm{10}_- \oplus \bm{5}^*_+ \oplus \bm{1}_+}
	&\oplus& \overbrace{\bm{5}_+ \oplus \bm{5}^*_-}
	&\oplus& \overbrace{\bm{1}_-} 
	\label{eq:matter-parity_2nd}
	\end{array} 
\end{align}
where subscript indicates matter-parity. Therefore, with the second assignment, matter superfields belong to $\bm{10}_- \in \bm{16}$, $\bm{5}^*_- \in \bm{10}$, $\bm{1}_- \in \bm{1}$, and the Higgs superfields belong to $\bm{5}^*_+ \in \bm{16}$ and $\bm{5}_+ \in \bm{10}$. 

The third assignment defined by eq.~\eqref{eq:B-L3} gives a different situation, since in this case matter-parity is not orthogonal to $SU(5)$. Therefore, eq.~\eqref{eq:matter-parity_1st} or eq.~\eqref{eq:matter-parity_2nd} becomes more complicated. In this case matter-parity can be written as follows,
\begin{align}
	\bm{16} = & (\bm{3}, \bm{2})_- \oplus (\bm{3}^*, \bm{1})_+ \oplus (\bm{1}, \bm{1})_+ \nonumber \\
	&\oplus (\bm{3}^*, \bm{1})_- \oplus (\bm{1}, \bm{2})_+ \oplus (\bm{1}, \bm{1})_-  \label{eq:16B-L3}\\
	\bm{10} = &(\bm{3}, \bm{1})_+  \oplus (\bm{1}, \bm{2})_- \oplus (\bm{3}^*, \bm{1})_- \oplus (\bm{1}, \bm{2})_+ \\
	\bm{1} = &(\bm{1}, \bm{1})_-, \label{eq:1B-L3}
\end{align}
where subscript is matter-parity, and two numbers in parentheses indicate $SU(3)_C$ and $SU(2)_L$, respectively. Equations~\eqref{eq:16B-L3}-\eqref{eq:1B-L3} show that with the third assignment, matter superfields with matter-parity of $-1$, and the Higgs superfields with matter-parity of $+1$, are unified in the same representation of $SU(5)$. 

As explicitly shown above, even in any assignments, matter-parity is not orthogonal to $E_6$. Therefore, matter superfields and the Higgs superfields can be unified in a single gauge multiplet $\bm{27}$ of $E_6$. 

Next we consider hypercharge. Since $E_6$ contains $U(1)_Y$, the following is satisfied,
\begin{align}
	U(1)_Z \otimes U(1)_V \otimes U(1)_{V^\prime} \supset U(1)_{Y}. 
\end{align}
There are three assignments that are consistent with the Standard Model~\cite{Harada:2003sb, Kawamura:2013rj},
\begin{align}
	1) & & \frac{Y}{2} &= Z, & & SU(5) \supset U(1)_Y \label{eq:Y1}\\
	2) & & \frac{Y}{2} &=-\frac{1}{5}(Z+V),  & & SU(5) \not\supset U(1)_Y \label{eq:Y2}\\
	3) & & \frac{Y}{2} &=-\frac{1}{20}(4Z - V - 5V^\prime), & & SO(10) \not\supset U(1)_Y. \label{eq:Y3}
\end{align}
The first assignment given by eq.~\eqref{eq:Y1} is the hypercharge of the original Georgi-Glashow $SU(5)$ model~\cite{Georgi:1974sy}, the second one given by eq.~\eqref{eq:Y2} is that of the flipped $SU(5)$ model~\cite{Barr:1981qv}, and the third one given by eq.~\eqref{eq:Y3} is that of the $E$-twisting $SU(5)$ model~\cite{Harada:2003sb} or of the flipped $SO(10)$ model~\cite{Bertolini:2010yz}. These three assignments are related by $SU(2)$, that is a subgroup of $SU(3)_R$~\cite{Harada:2003sb}. 

The first assignment defined by eq.~\eqref{eq:Y1} has been most extensively studied in the literature. In that case, three gauge coupling constants of the Standard Model gauge groups should be {\it exactly} unified at the grand unification scale, since $SU(3)_C$, $SU(2)_L$ and $U(1)_Y$ are contained in a single unified gauge group $SU(5)$. In contrast to the first assignment, however, with the second assignment defined by eq.~\eqref{eq:Y2} or the third assignment of eq.~\eqref{eq:Y3}, gauge coupling constants of $SU(3)_C$, $SU(2)_L$ and $U(1)_Y$ should be {\it approximately} unified at the grand unification scale, since $U(1)_Y$ is not a subgroup of $SU(5)$. We regard this issue as important, and therefore we are not interested in the rigorous unification for the present purpose---in this paper we consider one-loop renormalization group equations of the gauge coupling constants, and do not consider two-loop and threshold corrections.

Here we give the mention of $E_6$ exotic fields. To show an explicit particle embedding, as a typical example, we consider the most familiar assignment defined by eq.~\eqref{eq:B-L1} and eq.~\eqref{eq:Y1}. In this case, $\bm{16}$, $\bm{10}$ and $\bm{1}$ of $SO(10)$ in $\bm{27}$ of $E_6$ are decomposed as follows,
\begin{align}
	\bm{16} =& (\bm{3}, \bm{2})_{\frac{1}{6}, \frac{1}{3}}^- \oplus (\bm{3}^*, \bm{1})_{-\frac{2}{3}, -\frac{1}{3}}^- \oplus (\bm{1}, \bm{1})_{1,1}^- 
	\nonumber \\ 
	&\oplus (\bm{3}^*, \bm{1})_{\frac{1}{3},-\frac{1}{3}}^- \oplus (\bm{1}, \bm{2})_{-\frac{1}{2}, -1}^- \oplus (\bm{1}, \bm{1})_{0, 1}^- \\
	\bm{10} = &(\bm{3}, \bm{1})_{-\frac{1}{3}, -\frac{2}{3}}^+  \oplus (\bm{1}, \bm{2})_{\frac{1}{2}, 0}^+ \oplus (\bm{3}^*, \bm{1})_{\frac{1}{3},\frac{2}{3}}^+
	\oplus (\bm{1}, \bm{2})_{-\frac{1}{2},0}^+ \\
	\bm{1} = &(\bm{1}, \bm{1})_{0, 0}^+
\end{align}
where left-subscript is $Y/2$, right-subscript is $B-L$, superscript indicates matter-parity, and two numbers in parentheses are representations of $SU(3)_C$ and $SU(2)_L$, respectively. All the matter superfields---quarks, charged leptons, left-handed and right-handed neutrinos---are contained in $\bm{16}$ with matter-parity of $-1$. The up-type Higgs doublet $H_u$ is $(\bm{1}, \bm{2})_{\frac{1}{2}, 0}^+ \in \bm{10}$ with matter-parity of $+1$, and the down-type Higgs doublet $H_d$ is $(\bm{1}, \bm{2})_{-\frac{1}{2},0}^+ \in \bm{10}$ with matter-parity of $+1$. Two colored states $(\bm{3}, \bm{1})_{-\frac{1}{3}, -\frac{2}{3}}^+ \in \bm{10}$, and $(\bm{3}^*, \bm{1})_{\frac{1}{3},\frac{2}{3}}^+ \in \bm{10}$ are usually written by $D$ and $\overline{D}$, respectively. These states have $B-L=\mp 2/3$ as of di-quarks or leptoquarks and matter-parity of $+1$. These $D$ and $\overline{D}$ are referred as $E_6$ exotic fields. They contribute to the running of $SU(3)_C$ and $U(1)_Y$ gauge coupling constants, if they are relevant at low energy. 

\section{Gauge coupling unification}

One-loop renormalization group equation for the gauge coupling constants is given by
\begin{align}
	\frac{d\alpha_i}{dt} = \frac{b_i}{2\pi} \alpha_i^2 \quad (i=1, 2, 3)
\end{align}
where $\alpha_i \equiv g_i^2/4\pi$, $t=\ln \mu$ with $\mu$ the renormalization scale, and the $U(1)$ gauge coupling constant $g_1$ is normalized by $g_1 = \sqrt{5/3} g_Y$. The coefficients $b_3$, $b_2$ and $b_1$ are given by
\begin{align}
	 b_3 &=  -\frac{1}{3}\left(11 - 2n_{\rm s}\right)N + \frac{2}{3}\left(2 + n_{\rm s}\right)N_g
		+ \frac{1}{3}\left(1 + 2 n_{\rm s}\right)N_D \label{eq:b3}\\
	 b_2 &=  -\frac{1}{3}\left(11 - 2n_{\rm s}\right)N + \frac{2}{3}\left(2 + n_{\rm s}\right)N_g
		+  \frac{1}{6}\left(1 + 2n_{\rm s}\right)N_H
		\label{eq:b2}\\
	 b_1 &=  
	 	\frac{2}{3}\left(2 + n_{\rm s}\right)N_g
		+ \frac{1}{10}\left(1 + 2n_{\rm s}\right)N_H
		+ \frac{2}{15}\left(1 + 2 n_{\rm s}\right)N_D, \label{eq:b1}
\end{align}
where $N=3$ for $b_3$, $N=2$ for $b_2$, $N_g$ is the number of generations of quarks and leptons, $N_H$ is the number of Higgs doublets, $N_D$ is the number of pairs of $E_6$ exotic fields $D \oplus \overline{D}$. A parameter $n_s$ is assigned as $n_{\rm s}=+1$ for supersymmetric theories, and  $n_{\rm s}=0$ for nonsupersymmetric theories. In the following we first consider supersymmetric models in section~\ref{subsec1}. Nonsupersymmetric models will be studied in section~\ref{subsec2}.

\subsection{\label{subsec1}Supersymmetric cases}

In the minimal supersymmetric Standard Model (MSSM), there are two Higgs doublets $H_u$ and $H_d$, and no $E_6$ exotic pairs $D \oplus \overline{D}$. Therefore, by substituting parameters $n_{\rm s}=+1$, $N_g=3$, $N_H=2$ and $N_D=0$ into equations~\eqref{eq:b3}-\eqref{eq:b1}, $b_i$ are obtained as $(b_3, b_2, b_1)=(-3, 1, \frac{33}{5})$. If there are extra Higgs doublets and/or $E_6$ exotic fields $D \oplus \overline{D}$, they contribute to $b_i$, and then $b_i$ change as $b_i \rightarrow b_i + \Delta b_i$, 
\begin{align}
	\Delta b_3 &= \frac{1}{3}(1+2n_{\rm s})N_D  = N_D, \label{eq:delta_b3}\\
	\Delta b_2 &= \frac{1}{6}(1+2n_{\rm s})(N_H-2)   = \frac{1}{2}N_H - 1, \label{eq:delta_b2}\\
	\Delta b_1 &= \frac{1}{10}(1+2n_{\rm s})(N_H-2) + \frac{2}{15}(1+2n_{\rm s})N_D   \nonumber\\
	&= \frac{3}{10}N_H + \frac{2}{5}N_D - \frac{3}{5},
	\label{eq:delta_b1}
\end{align}
where $n_{\rm s}=+1$ is used. If $\Delta b_3$, $\Delta b_2$ and $\Delta b_1$ are equivalent, gauge coupling unification of MSSM is preserved. Therefore, two equations, $\Delta b_3 = \Delta b_2$ and $\Delta b_3 = \Delta b_1$, are obtained.  
However, $N_D$ and $N_H$ are not uniquely determined from these two equations, since one of them is redundant---eqs.~\eqref{eq:delta_b3}$\sim$~\eqref{eq:delta_b1} indicate that if $\Delta b_3$ equals to $\Delta b_2$, $\Delta b_1$ automatically equals to $\Delta b_3$ or $\Delta b_2$. The independent condition is
\begin{align}
	\Delta b_i = N_D = \frac{1}{2}N_H - 1. \label{eq:uc_susy}
\end{align}
As far as eq.~\eqref{eq:uc_susy} is satisfied, gauge coupling unification of MSSM is preserved. Therefore, there are various models in which gauge coupling unification is realized. Table~\ref{tab:table1} is a summary of models.

\begin{table}[h]
\caption{\label{tab:table1}
A catalog of supersymmetric models with gauge coupling unification in which there are extra Higgs doublets and $E_6$ exotic fields. MSSM, 2HGM, MHU are abbreviations for minimal supersymmetric Standard Model, two-Higgs-generation-model, and matter-Higgs-unification, respectively. In this table, $\Delta b_i$ is the extra contribution to $b_i$ of MSSM, $N_H$ is the number of Higgs doublets, $N_D$ is the number of pairs of $E_6$ exotic fields $D \oplus \overline{D}$, $b_3$, $b_2$ and $b_1$ are the coefficients of beta-function of the Standard Model gauge coupling constants. }
\begin{ruledtabular}
\begin{tabular}{ccccr}
\textrm{model}&
 $\Delta b_i$ & $N_H$ & $N_D$ & $b_3, b_2, b_1$\\
\colrule
\textrm{MSSM} &0 & 2 & 0 & $-3$, $1$, $33/5$\\
\textrm{2HGM} &1 & 4 & 1 & $-2$, $2$, $38/5$\\
\textrm{MHU} &2 & 6 & 2 & $-1$, $3$, $43/5$\\
\textrm{model A} &3 & 8 & 3 & $0$, $4$, $48/5$\\
\textrm{model B} &4 & 10 & 4 & $1$, $5$, $53/5$\\ 
\textrm{model C} &5 & 12 & 5 & $2$, $6$, $58/5$\\  
\end{tabular}
\end{ruledtabular}
\end{table}

It is known that in supersymmetric $E_6$ models with three generations of $\bm{27}$ at low energy, two extra $SU(2)$ doublets are required for gauge coupling unification~\cite{Dienes:1996du}. It corresponds to model A in table~\ref{tab:table1}. Models of this type have been studied as $E_6$SSM model~\cite{King:2007uj}, and recently as $\eta$ model in ref.~\cite{Cho:2016afr}. From a different view, extending the MSSM by adding $n_5$ pairs of $\bm{5} \oplus \bm{5}^*$ of $SU(5)$ multiplets has been studied~\cite{Martin:1997ns, Hisano:2012wq}. Model A in table~\ref{tab:table1} corresponds to $n_5=3$. A type of model B with $(b_3, b_2, b_1)=(1, 5, 53/5)$ has been very recently studied as $\eta_{\rm BKM}$ model in ref.~\cite{Cho:2016afr}. 

\begin{figure}[h]
			\begin{center}
				\includegraphics[width=85mm]{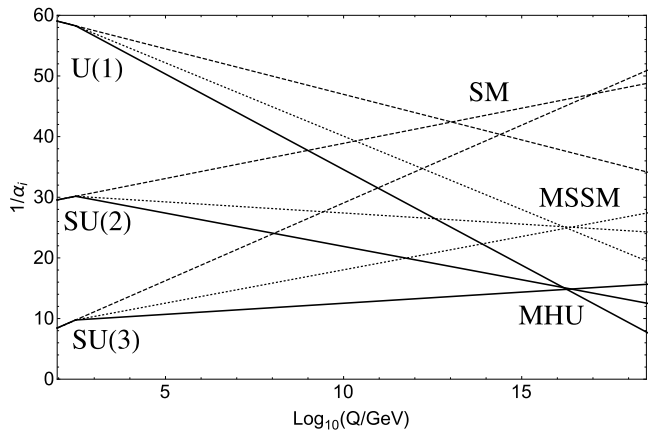}
				\caption{A figure of gauge coupling unification in the MHU model with supersymmetry (solid line). The gauge coupling running of the Standard Model (dashed line), and that of the minimal supersymmetric Standard Model (dotted line) are shown. }
				\label{fig:gaugeunification}
			\end{center} 
		\end{figure}

One of the attractive models in table~\ref{tab:table1} is the MHU model with $\Delta b_i = 2$, $N_H=6$ and $N_D=2$. In this model matter superfields and the doublets Higgs superfields are unified in a single gauge multiplet $\bm{27}$ of $E_6$, and then there are three Higgs generations. In the MHU model, the number of pairs of $E_6$ exotic fields $D\oplus \overline{D}$ is 2, and therefore one pair of $D \oplus \overline{D}$ in three generations should be decoupled. In this case, $b_3$, $b_2$ and $b_1$ are given by
\begin{align}
b_3 = -1, \quad b_2 = 3, \quad b_1 = \frac{43}{5}. 
\end{align}
Figure~\ref{fig:gaugeunification} shows the gauge coupling running with this coefficients (in the figure~\ref{fig:gaugeunification}, it is referred as MHU). For comparison, those of the Standard Model and minimal supersymmetric Standard Model are also shown.

Figure~\ref{fig:gaugeunification} indicates that in the MHU model the gauge coupling constant at the grand unification scale $M_{\rm GUT}$ is 
\begin{align}
	\alpha^{-1} (M_{\rm GUT}) \simeq 14, \qquad M_{\rm GUT} \simeq 2 \times 10^{16} {\rm GeV}. \label{eq:unifiedcoupling}
\end{align}
This is smaller than that of MSSM, $\alpha^{-1} (M_{\rm GUT}) \simeq 24$, and it enhances the decay rate of nucleus. The lifetime of the channel $p \rightarrow e^+ \pi^0$ is proportional to $\alpha^{-2}$. Therefore, by using the result in the minimal supersymmetric $SU(5)$ model~\cite{Hisano:2012wq}, the lifetime of proton can be estimated as
\begin{align}
	\tau /B(p \rightarrow e^+ \pi^0) \simeq 4 \times 10^{34} \times \left(\frac{M_X}{10^{16} {\rm GeV}}\right)^4{\rm years}, \label{eq:lifetime}
\end{align}
where $M_X$ is a mass of gauge boson in grand unified theories. Although this is roughly estimated value, it may be interesting---for a typical value of $M_X = 10^{16}$ GeV, eq.~\eqref{eq:lifetime} gives $\tau /B (p \rightarrow e^+ \pi^0) = 4 \times 10^{34}$ years that is just above the current SuperKamiokande limit $\tau /B(p \rightarrow e^+ \pi^0) = 1.29 \times 10^{34}$ years~\cite{Nishino:2012bnw}. 

\subsection{\label{subsec2}Nonsupersymmetric cases}

In this section we study nonsupersymmetric models. Although three gauge coupling constants are not unified in the Standard Model, we show that gauge coupling unification is realized if there are extra Higgs doublets and/or $E_6$ exotic fields $D \oplus \overline{D}$. 

In the Standard Model, by substituting $n_{\rm s}=0$, $N_g=3$, $N_H=1$ and $N_D=0$ into equations~\eqref{eq:b3}-\eqref{eq:b1}, $b_i$ are obtained as $(b_3, b_2, b_1)=(-7, -\frac{19}{6}, \frac{41}{10})$. Extra Higgs doublets and $E_6$ exotic fields $D \oplus \overline{D}$ contribute to $b_i$ of the Standard Model, and then the coefficient $b_i$ change as $b_i \rightarrow b_i + \delta b_i$, 
\begin{align}
	\delta b_3 &= \frac{1}{3}(1+2n_{\rm s})N_D  = \frac{1}{3}N_D, \label{eq:delta_b3_SM}\\
	\delta b_2 &= \frac{1}{6}(1+2n_{\rm s})(N_H-1)   = \frac{1}{6}(N_H - 1), \label{eq:delta_b2_SM}\\
	\delta b_1 &= \frac{1}{10}(1+2n_{\rm s})(N_H-1) + \frac{2}{15}(1+2n_{\rm s})N_D   \nonumber \\
	&= \frac{1}{10}(N_H-1) + \frac{2}{15}N_D,
	\label{eq:delta_b1_SM}	
\end{align}
where $n_{\rm s}=0$ is used. In this paper we count the number of $D \oplus \overline{D}$ as bosons. If they are counted as fermions, $N_D$ in eqs.~\eqref{eq:delta_b3_SM} and~\eqref{eq:delta_b1_SM} should be replaced by $2N_D$. 

\begin{figure}[h]
			\begin{center}
				\includegraphics[width=85mm]{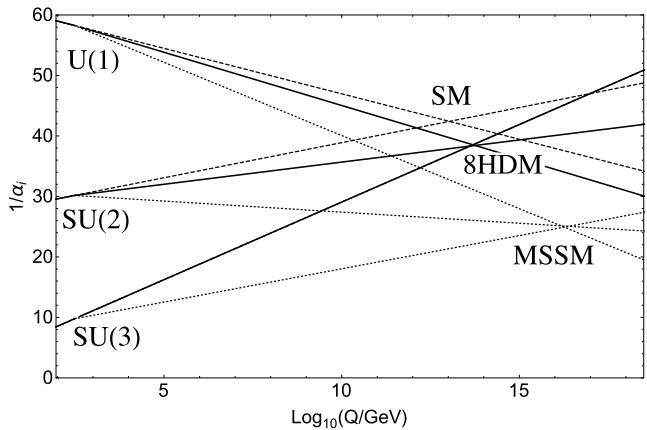}
				\caption{A figure of gauge coupling unification in eight-Higgs-doublet-model (8HDM) without supersymmetry (solid line). The gauge coupling running of the Standard Model (dashed line), and that of the minimal supersymmetric Standard Model (dotted line) are shown. }
				\label{fig:gaugeunification_nonsusy}
			\end{center} 
		\end{figure}

The nontrivial extension of the Standard Model is the model with $N_H=8$, $N_D = 0$ in which there are eight Higgs doublets and no $E_6$ exotic fields $D \oplus \overline{D}$. In this model, eqs.~\eqref{eq:delta_b3_SM}-\eqref{eq:delta_b1_SM} give $(\delta b_3, \delta b_2, \delta b_1) = (0, \frac{7}{6}, \frac{7}{10})$, and then $b_3$, $b_2$ and $b_1$ are given by
\begin{align}
	b_3=-7, \quad b_2=-2, \quad b_1=\frac{24}{5}.
\end{align}
Figure~\ref{fig:gaugeunification_nonsusy} shows the gauge coupling running in this model. In figure~\ref{fig:gaugeunification_nonsusy}, it is referred as eight-Higgs-doublet model (8HDM). 

Figure~\ref{fig:gaugeunification_nonsusy} indicates that even if nonsupersymmetric cases gauge coupling constants are unified at the grand unification scale $M_{\rm GUT}$, and the gauge coupling constant at $M_{\rm GUT}$ is given by
\begin{align}
	\alpha^{-1} (M_{\rm GUT}) \simeq 38, \qquad M_{\rm GUT} \simeq 5 \times 10^{13} {\rm GeV}.
\end{align}

Further extension is possible. If $\delta b_3$ equals to $\delta b_2 - \frac{7}{6}$ and $\delta b_1 - \frac{7}{10}$, the gauge coupling unification of 8HDM is preserved, and then following equations are obtained,
\begin{align}
	\frac{1}{3} N_D = \frac{1}{6} N_H - \frac{4}{3} = \frac{1}{10} N_H + \frac{2}{15} N_D - \frac{4}{5}.
\end{align}
The independent condition is the following,
\begin{align}
	N_D = \frac{1}{2} N_H - 4. \label{eq:uc_nonsusy}
\end{align}
As far as eq.~\eqref{eq:uc_nonsusy} is satisfied, the gauge coupling unification of 8HDM is preserved. Table~\ref{tab:table2} is a summary.

\begin{table}[h]
\caption{\label{tab:table2}
A list of nonsupersymmetric models in which there are extra Higgs doublets and $E_6$ exotic fields. In all models---except for the Standard Model---gauge coupling unification is realized. $N_H$ is the number of Higgs doublets, $N_D$ is the number of pairs of $E_6$ exotic fields $D \oplus \overline{D}$, $b_i \ (i=1, 2, 3)$ are the coefficients of beta-function of the Standard Model gauge coupling constants. 8HDM is an abbreviation for eight-Higgs-doublet-model.}
\begin{ruledtabular}
\begin{tabular}{cccr}
\textrm{model}&
 $N_H$ & $N_D$ & $b_3, b_2, b_1$\\
\colrule
\textrm{SM} & 1 & 0 & $-7$, $-19/6$, $41/10$\\
\textrm{8HDM} & 8 & 0 & $-7$, $-2$, $24/5$\\
\textrm{model D} & 10 & 1 & $-20/3$, $-5/3$, $77/15$\\
\textrm{model E} & 12 & 2 & $-19/3$, $-4/3$, $82/15$\\
\textrm{model F} & 14 & 3 & $-6$, $-1$, $29/5$\\ 
\textrm{model G} & 16 & 4 & $-17/3$, $-2/3$, $92/15$\\  
\end{tabular}
\end{ruledtabular}
\end{table}

\section{Summary}

In summary, we have studied the gauge coupling unification within the framework where there are extra Higgs doublets and/or $E_6$ exotic fields. We have presented a catalog of models in which gauge coupling unification is realized. The catalog covers both supersymmetric models (table~\ref{tab:table1}) and nonsupersymmetric models (table~\ref{tab:table2}).

\bibliography{references}

\begin{thebibliography}{38}%
\makeatletter
\providecommand \@ifxundefined [1]{%
 \@ifx{#1\undefined}
}%
\providecommand \@ifnum [1]{%
 \ifnum #1\expandafter \@firstoftwo
 \else \expandafter \@secondoftwo
 \fi
}%
\providecommand \@ifx [1]{%
 \ifx #1\expandafter \@firstoftwo
 \else \expandafter \@secondoftwo
 \fi
}%
\providecommand \natexlab [1]{#1}%
\providecommand \enquote  [1]{``#1''}%
\providecommand \bibnamefont  [1]{#1}%
\providecommand \bibfnamefont [1]{#1}%
\providecommand \citenamefont [1]{#1}%
\providecommand \href@noop [0]{\@secondoftwo}%
\providecommand \href [0]{\begingroup \@sanitize@url \@href}%
\providecommand \@href[1]{\@@startlink{#1}\@@href}%
\providecommand \@@href[1]{\endgroup#1\@@endlink}%
\providecommand \@sanitize@url [0]{\catcode `\\12\catcode `\$12\catcode
  `\&12\catcode `\#12\catcode `\^12\catcode `\_12\catcode `\%12\relax}%
\providecommand \@@startlink[1]{}%
\providecommand \@@endlink[0]{}%
\providecommand \url  [0]{\begingroup\@sanitize@url \@url }%
\providecommand \@url [1]{\endgroup\@href {#1}{\urlprefix }}%
\providecommand \urlprefix  [0]{URL }%
\providecommand \Eprint [0]{\href }%
\@ifxundefined \urlstyle {%
  \providecommand \doi  [0]{\begingroup \@sanitize@url \@doi}%
  \providecommand \@doi [1]{\endgroup \@@startlink {\doibase
  #1}doi:\discretionary {}{}{}#1\@@endlink }%
}{%
  \providecommand \doi  [0]{doi:\discretionary{}{}{}\begingroup
  \urlstyle{rm}\Url }%
}%
\providecommand \doibase [0]{http://dx.doi.org/}%
\providecommand \Doi [0]{\begingroup \@sanitize@url \@Doi }%
\providecommand \@Doi  [1]{\endgroup\@@startlink{\doibase#1}\@@Doi}%
\providecommand \@@Doi [1]{#1\@@endlink}%
\providecommand \selectlanguage [0]{\@gobble}%
\providecommand \bibinfo  [0]{\@secondoftwo}%
\providecommand \bibfield  [0]{\@secondoftwo}%
\providecommand \translation [1]{[#1]}%
\providecommand \BibitemOpen [0]{}%
\providecommand \bibitemStop [0]{}%
\providecommand \bibitemNoStop [0]{.\EOS\space}%
\providecommand \EOS [0]{\spacefactor3000\relax}%
\providecommand \BibitemShut  [1]{\csname bibitem#1\endcsname}%
\bibitem [{\citenamefont {Aad}\ \emph {et~al.}(2012)\citenamefont {Aad} \emph
  {et~al.}}]{Aad:2012tfa}%
  \BibitemOpen
  \bibfield  {author} {\bibinfo {author} {\bibfnamefont {G.}~\bibnamefont
  {Aad}} \emph {et~al.} (\bibinfo {collaboration} {ATLAS}),\ }\Doi
  {10.1016/j.physletb.2012.08.020} {\bibfield  {journal} {\bibinfo  {journal}
  {Phys. Lett.},\ }\textbf {\bibinfo {volume} {B716}},\ \bibinfo {pages} {1}
  (\bibinfo {year} {2012})},\ \Eprint {http://arxiv.org/abs/1207.7214}
  {arXiv:1207.7214 [hep-ex]} \BibitemShut {NoStop}%
\bibitem [{\citenamefont {Chatrchyan}\ \emph {et~al.}(2012)\citenamefont
  {Chatrchyan} \emph {et~al.}}]{Chatrchyan:2012xdj}%
  \BibitemOpen
  \bibfield  {author} {\bibinfo {author} {\bibfnamefont {S.}~\bibnamefont
  {Chatrchyan}} \emph {et~al.} (\bibinfo {collaboration} {CMS}),\ }\Doi
  {10.1016/j.physletb.2012.08.021} {\bibfield  {journal} {\bibinfo  {journal}
  {Phys. Lett.},\ }\textbf {\bibinfo {volume} {B716}},\ \bibinfo {pages} {30}
  (\bibinfo {year} {2012})},\ \Eprint {http://arxiv.org/abs/1207.7235}
  {arXiv:1207.7235 [hep-ex]} \BibitemShut {NoStop}%
\bibitem [{\citenamefont {Branco}\ \emph {et~al.}(2012)\citenamefont {Branco},
  \citenamefont {Ferreira}, \citenamefont {Lavoura}, \citenamefont {Rebelo},
  \citenamefont {Sher},\ and\ \citenamefont {Silva}}]{Branco:2011iw}%
  \BibitemOpen
  \bibfield  {author} {\bibinfo {author} {\bibfnamefont {G.~C.}\ \bibnamefont
  {Branco}}, \bibinfo {author} {\bibfnamefont {P.~M.}\ \bibnamefont
  {Ferreira}}, \bibinfo {author} {\bibfnamefont {L.}~\bibnamefont {Lavoura}},
  \bibinfo {author} {\bibfnamefont {M.~N.}\ \bibnamefont {Rebelo}}, \bibinfo
  {author} {\bibfnamefont {M.}~\bibnamefont {Sher}}, \ and\ \bibinfo {author}
  {\bibfnamefont {J.~P.}\ \bibnamefont {Silva}},\ }\Doi
  {10.1016/j.physrep.2012.02.002} {\bibfield  {journal} {\bibinfo  {journal}
  {Phys. Rept.},\ }\textbf {\bibinfo {volume} {516}},\ \bibinfo {pages} {1}
  (\bibinfo {year} {2012})},\ \Eprint {http://arxiv.org/abs/1106.0034}
  {arXiv:1106.0034 [hep-ph]} \BibitemShut {NoStop}%
\bibitem [{\citenamefont {Grossman}(1994)}]{Grossman:1994jb}%
  \BibitemOpen
  \bibfield  {author} {\bibinfo {author} {\bibfnamefont {Y.}~\bibnamefont
  {Grossman}},\ }\Doi {10.1016/0550-3213(94)90316-6} {\bibfield  {journal}
  {\bibinfo  {journal} {Nucl. Phys.},\ }\textbf {\bibinfo {volume} {B426}},\
  \bibinfo {pages} {355} (\bibinfo {year} {1994})},\ \Eprint
  {http://arxiv.org/abs/hep-ph/9401311} {arXiv:hep-ph/9401311 [hep-ph]}
  \BibitemShut {NoStop}%
\bibitem [{\citenamefont {Kilian}\ and\ \citenamefont
  {Reuter}(2006)}]{Kilian:2006hh}%
  \BibitemOpen
  \bibfield  {author} {\bibinfo {author} {\bibfnamefont {W.}~\bibnamefont
  {Kilian}}\ and\ \bibinfo {author} {\bibfnamefont {J.}~\bibnamefont
  {Reuter}},\ }\Doi {10.1016/j.physletb.2006.09.033} {\bibfield  {journal}
  {\bibinfo  {journal} {Phys. Lett.},\ }\textbf {\bibinfo {volume} {B642}},\
  \bibinfo {pages} {81} (\bibinfo {year} {2006})},\ \Eprint
  {http://arxiv.org/abs/hep-ph/0606277} {arXiv:hep-ph/0606277 [hep-ph]}
  \BibitemShut {NoStop}%
\bibitem [{\citenamefont {Hartmann}\ and\ \citenamefont
  {Kilian}(2014)}]{Hartmann:2014ppa}%
  \BibitemOpen
  \bibfield  {author} {\bibinfo {author} {\bibfnamefont {F.}~\bibnamefont
  {Hartmann}}\ and\ \bibinfo {author} {\bibfnamefont {W.}~\bibnamefont
  {Kilian}},\ }\Doi {10.1140/epjc/s10052-014-3055-4} {\bibfield  {journal}
  {\bibinfo  {journal} {Eur. Phys. J.},\ }\textbf {\bibinfo {volume} {C74}},\
  \bibinfo {pages} {3055} (\bibinfo {year} {2014})},\ \Eprint
  {http://arxiv.org/abs/1405.1901} {arXiv:1405.1901 [hep-ph]} \BibitemShut
  {NoStop}%
\bibitem [{\citenamefont {King}\ \emph {et~al.}(2007)\citenamefont {King},
  \citenamefont {Moretti},\ and\ \citenamefont {Nevzorov}}]{King:2007uj}%
  \BibitemOpen
  \bibfield  {author} {\bibinfo {author} {\bibfnamefont {S.~F.}\ \bibnamefont
  {King}}, \bibinfo {author} {\bibfnamefont {S.}~\bibnamefont {Moretti}}, \
  and\ \bibinfo {author} {\bibfnamefont {R.}~\bibnamefont {Nevzorov}},\ }\Doi
  {10.1016/j.physletb.2007.04.061} {\bibfield  {journal} {\bibinfo  {journal}
  {Phys. Lett.},\ }\textbf {\bibinfo {volume} {B650}},\ \bibinfo {pages} {57}
  (\bibinfo {year} {2007})},\ \Eprint {http://arxiv.org/abs/hep-ph/0701064}
  {arXiv:hep-ph/0701064 [hep-ph]} \BibitemShut {NoStop}%
\bibitem [{\citenamefont {Hartmann}\ \emph {et~al.}(2014)\citenamefont
  {Hartmann}, \citenamefont {Kilian},\ and\ \citenamefont
  {Schnitter}}]{Hartmann:2014fya}%
  \BibitemOpen
  \bibfield  {author} {\bibinfo {author} {\bibfnamefont {F.}~\bibnamefont
  {Hartmann}}, \bibinfo {author} {\bibfnamefont {W.}~\bibnamefont {Kilian}}, \
  and\ \bibinfo {author} {\bibfnamefont {K.}~\bibnamefont {Schnitter}},\ }\Doi
  {10.1007/JHEP05(2014)064} {\bibfield  {journal} {\bibinfo  {journal} {JHEP},\
  }\textbf {\bibinfo {volume} {05}},\ \bibinfo {pages} {064} (\bibinfo {year}
  {2014})},\ \Eprint {http://arxiv.org/abs/1401.7891} {arXiv:1401.7891
  [hep-ph]} \BibitemShut {NoStop}%
\bibitem [{\citenamefont {Cho}\ \emph {et~al.}()\citenamefont {Cho},
  \citenamefont {Maru},\ and\ \citenamefont {Yotsutani}}]{Cho:2016afr}%
  \BibitemOpen
  \bibfield  {author} {\bibinfo {author} {\bibfnamefont {G.-C.}\ \bibnamefont
  {Cho}}, \bibinfo {author} {\bibfnamefont {N.}~\bibnamefont {Maru}}, \ and\
  \bibinfo {author} {\bibfnamefont {K.}~\bibnamefont {Yotsutani}},\ }\href@noop
  {} {}\Eprint {http://arxiv.org/abs/1602.04271} {arXiv:1602.04271 [hep-ph]}
  \BibitemShut {NoStop}%
\bibitem [{\citenamefont {Nishino}\ \emph {et~al.}(2012)\citenamefont {Nishino}
  \emph {et~al.}}]{Nishino:2012bnw}%
  \BibitemOpen
  \bibfield  {author} {\bibinfo {author} {\bibfnamefont {H.}~\bibnamefont
  {Nishino}} \emph {et~al.} (\bibinfo {collaboration} {Super-Kamiokande}),\
  }\Doi {10.1103/PhysRevD.85.112001} {\bibfield  {journal} {\bibinfo  {journal}
  {Phys. Rev.},\ }\textbf {\bibinfo {volume} {D85}},\ \bibinfo {pages} {112001}
  (\bibinfo {year} {2012})},\ \Eprint {http://arxiv.org/abs/1203.4030}
  {arXiv:1203.4030 [hep-ex]} \BibitemShut {NoStop}%
\bibitem [{\citenamefont {Gursey}\ \emph {et~al.}(1976)\citenamefont {Gursey},
  \citenamefont {Ramond},\ and\ \citenamefont {Sikivie}}]{Gursey:1975ki}%
  \BibitemOpen
  \bibfield  {author} {\bibinfo {author} {\bibfnamefont {F.}~\bibnamefont
  {Gursey}}, \bibinfo {author} {\bibfnamefont {P.}~\bibnamefont {Ramond}}, \
  and\ \bibinfo {author} {\bibfnamefont {P.}~\bibnamefont {Sikivie}},\ }\Doi
  {10.1016/0370-2693(76)90417-2} {\bibfield  {journal} {\bibinfo  {journal}
  {Phys. Lett.},\ }\textbf {\bibinfo {volume} {B60}},\ \bibinfo {pages} {177}
  (\bibinfo {year} {1976})}\BibitemShut {NoStop}%
\bibitem [{\citenamefont {Achiman}\ and\ \citenamefont
  {Stech}(1978)}]{Achiman:1978vg}%
  \BibitemOpen
  \bibfield  {author} {\bibinfo {author} {\bibfnamefont {Y.}~\bibnamefont
  {Achiman}}\ and\ \bibinfo {author} {\bibfnamefont {B.}~\bibnamefont
  {Stech}},\ }\Doi {10.1016/0370-2693(78)90584-1} {\bibfield  {journal}
  {\bibinfo  {journal} {Phys. Lett.},\ }\textbf {\bibinfo {volume} {B77}},\
  \bibinfo {pages} {389} (\bibinfo {year} {1978})}\BibitemShut {NoStop}%
\bibitem [{\citenamefont {Ruegg}\ and\ \citenamefont
  {Schucker}(1979)}]{Ruegg:1979fr}%
  \BibitemOpen
  \bibfield  {author} {\bibinfo {author} {\bibfnamefont {H.}~\bibnamefont
  {Ruegg}}\ and\ \bibinfo {author} {\bibfnamefont {T.}~\bibnamefont
  {Schucker}},\ }\Doi {10.1016/0550-3213(79)90219-0} {\bibfield  {journal}
  {\bibinfo  {journal} {Nucl. Phys.},\ }\textbf {\bibinfo {volume} {B161}},\
  \bibinfo {pages} {388} (\bibinfo {year} {1979})}\BibitemShut {NoStop}%
\bibitem [{\citenamefont {Barbieri}\ \emph {et~al.}(1981)\citenamefont
  {Barbieri}, \citenamefont {Nanopoulos},\ and\ \citenamefont
  {Masiero}}]{Barbieri:1981yy}%
  \BibitemOpen
  \bibfield  {author} {\bibinfo {author} {\bibfnamefont {R.}~\bibnamefont
  {Barbieri}}, \bibinfo {author} {\bibfnamefont {D.~V.}\ \bibnamefont
  {Nanopoulos}}, \ and\ \bibinfo {author} {\bibfnamefont {A.}~\bibnamefont
  {Masiero}},\ }\Doi {10.1016/0370-2693(81)90589-X} {\bibfield  {journal}
  {\bibinfo  {journal} {Phys. Lett.},\ }\textbf {\bibinfo {volume} {B104}},\
  \bibinfo {pages} {194} (\bibinfo {year} {1981})}\BibitemShut {NoStop}%
\bibitem [{\citenamefont {Gursey}\ and\ \citenamefont
  {Serdaroglu}(1981)}]{Gursey:1981kf}%
  \BibitemOpen
  \bibfield  {author} {\bibinfo {author} {\bibfnamefont {F.}~\bibnamefont
  {Gursey}}\ and\ \bibinfo {author} {\bibfnamefont {M.}~\bibnamefont
  {Serdaroglu}},\ }\Doi {10.1007/BF02827439} {\bibfield  {journal} {\bibinfo
  {journal} {Nuovo Cim.},\ }\textbf {\bibinfo {volume} {A65}},\ \bibinfo
  {pages} {337} (\bibinfo {year} {1981})}\BibitemShut {NoStop}%
\bibitem [{\citenamefont {Robinett}(1982)}]{Robinett:1982gw}%
  \BibitemOpen
  \bibfield  {author} {\bibinfo {author} {\bibfnamefont {R.~W.}\ \bibnamefont
  {Robinett}},\ }\Doi {10.1103/PhysRevD.26.2388} {\bibfield  {journal}
  {\bibinfo  {journal} {Phys. Rev.},\ }\textbf {\bibinfo {volume} {D26}},\
  \bibinfo {pages} {2388} (\bibinfo {year} {1982})}\BibitemShut {NoStop}%
\bibitem [{\citenamefont {Sen}(1985)}]{Sen:1985af}%
  \BibitemOpen
  \bibfield  {author} {\bibinfo {author} {\bibfnamefont {A.}~\bibnamefont
  {Sen}},\ }\Doi {10.1103/PhysRevLett.55.33} {\bibfield  {journal} {\bibinfo
  {journal} {Phys. Rev. Lett.},\ }\textbf {\bibinfo {volume} {55}},\ \bibinfo
  {pages} {33} (\bibinfo {year} {1985})}\BibitemShut {NoStop}%
\bibitem [{\citenamefont {Robinett}(1986)}]{Robinett:1985dz}%
  \BibitemOpen
  \bibfield  {author} {\bibinfo {author} {\bibfnamefont {R.~W.}\ \bibnamefont
  {Robinett}},\ }\Doi {10.1103/PhysRevD.33.1908} {\bibfield  {journal}
  {\bibinfo  {journal} {Phys. Rev.},\ }\textbf {\bibinfo {volume} {D33}},\
  \bibinfo {pages} {1908} (\bibinfo {year} {1986})}\BibitemShut {NoStop}%
\bibitem [{\citenamefont {Bando}\ and\ \citenamefont
  {Kugo}(1999)}]{Bando:1999km}%
  \BibitemOpen
  \bibfield  {author} {\bibinfo {author} {\bibfnamefont {M.}~\bibnamefont
  {Bando}}\ and\ \bibinfo {author} {\bibfnamefont {T.}~\bibnamefont {Kugo}},\
  }\Doi {10.1143/PTP.101.1313} {\bibfield  {journal} {\bibinfo  {journal}
  {Prog. Theor. Phys.},\ }\textbf {\bibinfo {volume} {101}},\ \bibinfo {pages}
  {1313} (\bibinfo {year} {1999})},\ \Eprint
  {http://arxiv.org/abs/hep-ph/9902204} {arXiv:hep-ph/9902204 [hep-ph]}
  \BibitemShut {NoStop}%
\bibitem [{\citenamefont {Bando}\ \emph {et~al.}(2000)\citenamefont {Bando},
  \citenamefont {Kugo},\ and\ \citenamefont {Yoshioka}}]{Bando:2000gs}%
  \BibitemOpen
  \bibfield  {author} {\bibinfo {author} {\bibfnamefont {M.}~\bibnamefont
  {Bando}}, \bibinfo {author} {\bibfnamefont {T.}~\bibnamefont {Kugo}}, \ and\
  \bibinfo {author} {\bibfnamefont {K.}~\bibnamefont {Yoshioka}},\ }\Doi
  {10.1143/PTP.104.211} {\bibfield  {journal} {\bibinfo  {journal} {Prog.
  Theor. Phys.},\ }\textbf {\bibinfo {volume} {104}},\ \bibinfo {pages} {211}
  (\bibinfo {year} {2000})},\ \Eprint {http://arxiv.org/abs/hep-ph/0003220}
  {arXiv:hep-ph/0003220 [hep-ph]} \BibitemShut {NoStop}%
\bibitem [{\citenamefont {Harada}(2003)}]{Harada:2003sb}%
  \BibitemOpen
  \bibfield  {author} {\bibinfo {author} {\bibfnamefont {J.}~\bibnamefont
  {Harada}},\ }\Doi {10.1088/1126-6708/2003/04/011} {\bibfield  {journal}
  {\bibinfo  {journal} {JHEP},\ }\textbf {\bibinfo {volume} {04}},\ \bibinfo
  {pages} {011} (\bibinfo {year} {2003})},\ \Eprint
  {http://arxiv.org/abs/hep-ph/0305015} {arXiv:hep-ph/0305015 [hep-ph]}
  \BibitemShut {NoStop}%
\bibitem [{\citenamefont {Maekawa}\ and\ \citenamefont
  {Yamashita}(2003)}]{Maekawa:2003ka}%
  \BibitemOpen
  \bibfield  {author} {\bibinfo {author} {\bibfnamefont {N.}~\bibnamefont
  {Maekawa}}\ and\ \bibinfo {author} {\bibfnamefont {T.}~\bibnamefont
  {Yamashita}},\ }\Doi {10.1103/PhysRevD.68.055001} {\bibfield  {journal}
  {\bibinfo  {journal} {Phys. Rev.},\ }\textbf {\bibinfo {volume} {D68}},\
  \bibinfo {pages} {055001} (\bibinfo {year} {2003})},\ \Eprint
  {http://arxiv.org/abs/hep-ph/0305116} {arXiv:hep-ph/0305116 [hep-ph]}
  \BibitemShut {NoStop}%
\bibitem [{\citenamefont {Stech}\ and\ \citenamefont
  {Tavartkiladze}(2004)}]{Stech:2003sb}%
  \BibitemOpen
  \bibfield  {author} {\bibinfo {author} {\bibfnamefont {B.}~\bibnamefont
  {Stech}}\ and\ \bibinfo {author} {\bibfnamefont {Z.}~\bibnamefont
  {Tavartkiladze}},\ }\Doi {10.1103/PhysRevD.70.035002} {\bibfield  {journal}
  {\bibinfo  {journal} {Phys. Rev.},\ }\textbf {\bibinfo {volume} {D70}},\
  \bibinfo {pages} {035002} (\bibinfo {year} {2004})},\ \Eprint
  {http://arxiv.org/abs/hep-ph/0311161} {arXiv:hep-ph/0311161 [hep-ph]}
  \BibitemShut {NoStop}%
\bibitem [{\citenamefont {Frank}\ \emph {et~al.}(2005)\citenamefont {Frank},
  \citenamefont {Turan},\ and\ \citenamefont {Sher}}]{Frank:2004vg}%
  \BibitemOpen
  \bibfield  {author} {\bibinfo {author} {\bibfnamefont {M.}~\bibnamefont
  {Frank}}, \bibinfo {author} {\bibfnamefont {I.}~\bibnamefont {Turan}}, \ and\
  \bibinfo {author} {\bibfnamefont {M.}~\bibnamefont {Sher}},\ }\Doi
  {10.1103/PhysRevD.71.113001} {\bibfield  {journal} {\bibinfo  {journal}
  {Phys. Rev.},\ }\textbf {\bibinfo {volume} {D71}},\ \bibinfo {pages} {113001}
  (\bibinfo {year} {2005})},\ \Eprint {http://arxiv.org/abs/hep-ph/0412090}
  {arXiv:hep-ph/0412090 [hep-ph]} \BibitemShut {NoStop}%
\bibitem [{\citenamefont {Stech}\ and\ \citenamefont
  {Tavartkiladze}(2008)}]{Stech:2008wd}%
  \BibitemOpen
  \bibfield  {author} {\bibinfo {author} {\bibfnamefont {B.}~\bibnamefont
  {Stech}}\ and\ \bibinfo {author} {\bibfnamefont {Z.}~\bibnamefont
  {Tavartkiladze}},\ }\Doi {10.1103/PhysRevD.77.076009} {\bibfield  {journal}
  {\bibinfo  {journal} {Phys. Rev.},\ }\textbf {\bibinfo {volume} {D77}},\
  \bibinfo {pages} {076009} (\bibinfo {year} {2008})},\ \Eprint
  {http://arxiv.org/abs/0802.0894} {arXiv:0802.0894 [hep-ph]} \BibitemShut
  {NoStop}%
\bibitem [{\citenamefont {Kawase}\ and\ \citenamefont
  {Maekawa}(2010)}]{Kawase:2010na}%
  \BibitemOpen
  \bibfield  {author} {\bibinfo {author} {\bibfnamefont {H.}~\bibnamefont
  {Kawase}}\ and\ \bibinfo {author} {\bibfnamefont {N.}~\bibnamefont
  {Maekawa}},\ }\Doi {10.1143/PTP.123.941} {\bibfield  {journal} {\bibinfo
  {journal} {Prog. Theor. Phys.},\ }\textbf {\bibinfo {volume} {123}},\
  \bibinfo {pages} {941} (\bibinfo {year} {2010})},\ \Eprint
  {http://arxiv.org/abs/1005.1049} {arXiv:1005.1049 [hep-ph]} \BibitemShut
  {NoStop}%
\bibitem [{\citenamefont {Stech}(2010)}]{Stech:2010tv}%
  \BibitemOpen
  \bibfield  {author} {\bibinfo {author} {\bibfnamefont {B.}~\bibnamefont
  {Stech}},\ }\bibfield  {booktitle} {\emph {\bibinfo {booktitle} {{Elementary
  particle physics and gravity. Proceedings, Corfu Summer Institute, School and
  Workshops on 'Standard model and beyond and standard cosmology' and on
  'Cosmology and strings: Theory - cosmology - phenomenology', CORFU 2009,
  Corfu, Greece, August 30-September 13, 2009}}},\ }\Doi
  {10.1002/prop.201000034} {\bibfield  {journal} {\bibinfo  {journal} {Fortsch.
  Phys.},\ }\textbf {\bibinfo {volume} {58}},\ \bibinfo {pages} {692} (\bibinfo
  {year} {2010})},\ \Eprint {http://arxiv.org/abs/1003.0581} {arXiv:1003.0581
  [hep-ph]} \BibitemShut {NoStop}%
\bibitem [{\citenamefont {Chen}\ and\ \citenamefont
  {Chung}(2011)}]{Chen:2010tg}%
  \BibitemOpen
  \bibfield  {author} {\bibinfo {author} {\bibfnamefont {C.-M.}\ \bibnamefont
  {Chen}}\ and\ \bibinfo {author} {\bibfnamefont {Y.-C.}\ \bibnamefont
  {Chung}},\ }\Doi {10.1007/JHEP03(2011)129} {\bibfield  {journal} {\bibinfo
  {journal} {JHEP},\ }\textbf {\bibinfo {volume} {03}},\ \bibinfo {pages} {129}
  (\bibinfo {year} {2011})},\ \Eprint {http://arxiv.org/abs/1010.5536}
  {arXiv:1010.5536 [hep-th]} \BibitemShut {NoStop}%
\bibitem [{\citenamefont {Di~Luzio}(2011)}]{DiLuzio:2011my}%
  \BibitemOpen
  \bibfield  {author} {\bibinfo {author} {\bibfnamefont {L.}~\bibnamefont
  {Di~Luzio}},\ }\emph {\bibinfo {title} {{Aspects of symmetry breaking in
  Grand Unified Theories}}},\ \href
  {http://inspirehep.net/record/940028/files/arXiv:1110.3210.pdf} {Ph.D.
  thesis},\ \bibinfo  {school} {SISSA, Trieste} (\bibinfo {year} {2011}),\
  \Eprint {http://arxiv.org/abs/1110.3210} {arXiv:1110.3210 [hep-ph]}
  \BibitemShut {NoStop}%
\bibitem [{\citenamefont {Kawamura}\ and\ \citenamefont
  {Miura}(2013)}]{Kawamura:2013rj}%
  \BibitemOpen
  \bibfield  {author} {\bibinfo {author} {\bibfnamefont {Y.}~\bibnamefont
  {Kawamura}}\ and\ \bibinfo {author} {\bibfnamefont {T.}~\bibnamefont
  {Miura}},\ }\Doi {10.1142/S0217751X13500553} {\bibfield  {journal} {\bibinfo
  {journal} {Int. J. Mod. Phys.},\ }\textbf {\bibinfo {volume} {A28}},\
  \bibinfo {pages} {1350055} (\bibinfo {year} {2013})},\ \Eprint
  {http://arxiv.org/abs/1301.7469} {arXiv:1301.7469 [hep-ph]} \BibitemShut
  {NoStop}%
\bibitem [{\citenamefont {Huang}(2014)}]{Huang:2014zba}%
  \BibitemOpen
  \bibfield  {author} {\bibinfo {author} {\bibfnamefont {C.-S.}\ \bibnamefont
  {Huang}},\ }\Doi {10.1142/S0217732314501508} {\bibfield  {journal} {\bibinfo
  {journal} {Mod. Phys. Lett.},\ }\textbf {\bibinfo {volume} {A29}},\ \bibinfo
  {pages} {1450150} (\bibinfo {year} {2014})},\ \Eprint
  {http://arxiv.org/abs/1402.2737} {arXiv:1402.2737 [hep-ph]} \BibitemShut
  {NoStop}%
\bibitem [{\citenamefont {Drissi El~Bouzaidi}\ and\ \citenamefont
  {Nassiri}(2015)}]{and:2015uya}%
  \BibitemOpen
  \bibfield  {author} {\bibinfo {author} {\bibfnamefont {M.}~\bibnamefont
  {Drissi El~Bouzaidi}}\ and\ \bibinfo {author} {\bibfnamefont
  {S.}~\bibnamefont {Nassiri}},\ }\Doi {10.1002/prop.201500030} {\bibfield
  {journal} {\bibinfo  {journal} {Fortsch. Phys.},\ }\textbf {\bibinfo {volume}
  {63}},\ \bibinfo {pages} {596} (\bibinfo {year} {2015})}\BibitemShut
  {NoStop}%
\bibitem [{\citenamefont {Georgi}\ and\ \citenamefont
  {Glashow}(1974)}]{Georgi:1974sy}%
  \BibitemOpen
  \bibfield  {author} {\bibinfo {author} {\bibfnamefont {H.}~\bibnamefont
  {Georgi}}\ and\ \bibinfo {author} {\bibfnamefont {S.~L.}\ \bibnamefont
  {Glashow}},\ }\Doi {10.1103/PhysRevLett.32.438} {\bibfield  {journal}
  {\bibinfo  {journal} {Phys. Rev. Lett.},\ }\textbf {\bibinfo {volume} {32}},\
  \bibinfo {pages} {438} (\bibinfo {year} {1974})}\BibitemShut {NoStop}%
\bibitem [{\citenamefont {Barr}(1982)}]{Barr:1981qv}%
  \BibitemOpen
  \bibfield  {author} {\bibinfo {author} {\bibfnamefont {S.~M.}\ \bibnamefont
  {Barr}},\ }\Doi {10.1016/0370-2693(82)90966-2} {\bibfield  {journal}
  {\bibinfo  {journal} {Phys. Lett.},\ }\textbf {\bibinfo {volume} {B112}},\
  \bibinfo {pages} {219} (\bibinfo {year} {1982})}\BibitemShut {NoStop}%
\bibitem [{\citenamefont {Bertolini}\ \emph {et~al.}(2011)\citenamefont
  {Bertolini}, \citenamefont {Di~Luzio},\ and\ \citenamefont
  {Malinsky}}]{Bertolini:2010yz}%
  \BibitemOpen
  \bibfield  {author} {\bibinfo {author} {\bibfnamefont {S.}~\bibnamefont
  {Bertolini}}, \bibinfo {author} {\bibfnamefont {L.}~\bibnamefont {Di~Luzio}},
  \ and\ \bibinfo {author} {\bibfnamefont {M.}~\bibnamefont {Malinsky}},\ }\Doi
  {10.1103/PhysRevD.83.035002} {\bibfield  {journal} {\bibinfo  {journal}
  {Phys. Rev.},\ }\textbf {\bibinfo {volume} {D83}},\ \bibinfo {pages} {035002}
  (\bibinfo {year} {2011})},\ \Eprint {http://arxiv.org/abs/1011.1821}
  {arXiv:1011.1821 [hep-ph]} \BibitemShut {NoStop}%
\bibitem [{\citenamefont {Dienes}(1997)}]{Dienes:1996du}%
  \BibitemOpen
  \bibfield  {author} {\bibinfo {author} {\bibfnamefont {K.~R.}\ \bibnamefont
  {Dienes}},\ }\bibfield  {booktitle} {\emph {\bibinfo {booktitle} {{Institute
  for Theoretical Physics Conference on Unification: From the Weak Scale to the
  Planck Scale Santa Barbara, California, October 23-27, 1995}}},\ }\Doi
  {10.1016/S0370-1573(97)00009-4} {\bibfield  {journal} {\bibinfo  {journal}
  {Phys. Rept.},\ }\textbf {\bibinfo {volume} {287}},\ \bibinfo {pages} {447}
  (\bibinfo {year} {1997})},\ \Eprint {http://arxiv.org/abs/hep-th/9602045}
  {arXiv:hep-th/9602045 [hep-th]} \BibitemShut {NoStop}%
\bibitem [{\citenamefont {Martin}(1997)}]{Martin:1997ns}%
  \BibitemOpen
  \bibfield  {author} {\bibinfo {author} {\bibfnamefont {S.~P.}\ \bibnamefont
  {Martin}},\ }\Doi {10.1142/9789812839657_0001, 10.1142/9789814307505_0001} {
  (\bibinfo {year} {1997})},\ \doi {10.1142/9789812839657_0001,
  10.1142/9789814307505_0001},\ \bibinfo {note} {[Adv. Ser. Direct. High Energy
  Phys.18,1(1998)]},\ \Eprint {http://arxiv.org/abs/hep-ph/9709356}
  {arXiv:hep-ph/9709356 [hep-ph]} \BibitemShut {NoStop}%
\bibitem [{\citenamefont {Hisano}\ \emph {et~al.}(2012)\citenamefont {Hisano},
  \citenamefont {Kobayashi},\ and\ \citenamefont {Nagata}}]{Hisano:2012wq}%
  \BibitemOpen
  \bibfield  {author} {\bibinfo {author} {\bibfnamefont {J.}~\bibnamefont
  {Hisano}}, \bibinfo {author} {\bibfnamefont {D.}~\bibnamefont {Kobayashi}}, \
  and\ \bibinfo {author} {\bibfnamefont {N.}~\bibnamefont {Nagata}},\ }\Doi
  {10.1016/j.physletb.2012.08.037} {\bibfield  {journal} {\bibinfo  {journal}
  {Phys. Lett.},\ }\textbf {\bibinfo {volume} {B716}},\ \bibinfo {pages} {406}
  (\bibinfo {year} {2012})},\ \Eprint {http://arxiv.org/abs/1204.6274}
  {arXiv:1204.6274 [hep-ph]} \BibitemShut {NoStop}%
\end{thebibliography}%

\end{document}